\documentstyle[newarcrc,fleqn,graphicx]{article}

\def\VEL{\:{\rm km\:s^{-1}}}

\def\LA{Lyman\thinspace$\alpha$}

\def\sun{\hbox{$\odot$}}
\newcommand{\MSOL}{\mbox{$\:M_{\sun}$}}

\newcommand{\EXPU}[3]{\mbox{\rm $#1 \times 10^{#2} \rm\:#3$}}  

\title{What We Learn from Quantitative Ultraviolet Spectroscopy of Naked White Dwarfs in Cataclysmic Variables}

\author{Knox S. Long\address{Space Telescope Science Institute\\
3700 San Martin Drive\\Baltimore, MD 21218, United States
}
}

\begin{document}
\maketitle

\begin{abstract}
Using the Hopkins Ultraviolet Telescope and Hubble Space Telescope,
observers have now obtained UV spectra with sufficient signal
to noise and resolution to allow quantitative spectroscopic analyses of
the WDs in several DNe.  In the ``cleanest'' DNe,
such as U Gem, the observations are allowing the basic physical
parameters of the WD  -- temperature, radius, gravity, rotation rate,
and surface abundances -- to be established. 
A second component also exists in these systems,
which may either be the disk or may be related to the WD itself.  Here
I summarize the current state of the observations and
our understanding of the data, highlighting some of the uncertainties
in the analyses as well the prospects for fundamentally advancing our
understanding of DNe and WDs with future observations.
\end{abstract}

\section{Introduction}
White dwarfs in cataclysmic variables are important since they are the 
gravitational engines which power CVs.  Furthermore,
compared to most WDs, conditions on the surface of the WDs in CVs are extreme,
making these WDs basic laboratories for WD physics.  
With sensitive UV spectrographs in space, it has become possible to carry 
out detailed UV spectroscopy of WDs in some DNe.

The first UV spectra of WDs in CVs were obtained with IUE.  
Panek \& Holm [1985] showed that for U Gem in quiescence 
the depth of the \LA, the shape of the UV spectrum, and the overall flux 
were consistent with 
a 1.2 \MSOL WD at a distance of 90 pc with a temperature of 30,000 K. 
Mateo \& Szkody [1984] interpreted the deep \LA\ profile of VW Hyi as arising from a 20,000 K WD.  
Similarly, Holm [1988] argued that the spectrum of WZ Sge far from outburst resembled a DA WD.  
Some other systems, such as EK TrA and ST L Mi, and recently SW UMa and CU Vel 
\cite{GK99}, also 
show \LA\ absorption line profiles. However, in these systems it is less clear 
that the WD was dominates the spectrum, since 
models of optically thick accretion disks also have prominent \LA\ profiles.  
Furthermore, the IUE observations also showed that the majority of DNe have spectra, even in quiescence, that exhibit none of the signatures expected of a WD.

There are various ways to observe WDs at UV wavelengths in CVs.  
In eclipsing DNe such as OY Car, the WD can 
sometimes be observed as a sharp jump in the flux due to occultation of the 
WD by the secondary star.  From the magnitude of the flux change and the shape of the spectrum in such eclipsing systems, one can estimate the WD temperature.
On the other hand, a detailed analysis of the photometric spectrum is usually impossible, 
because of the effects of the so-called ``Fe-curtain'' on the WD spectrum \cite{horne94}.  And 
in polars like AM Her, the UV continuum, at least in some orbital phases, is  
likely dominated by  emission from the WD photosphere \cite{gansicke_amher}.  
However, detailed analysis is again difficult because radiation from the accretion column heats and alters the structure of the the photosphere.  
Here we will concentrate on WD in systems in which the WD dominates the UV emission, i.e. we will concentrate on ``naked WDs'' in DNe.

\section{U Geminorum}

A qualitative improvement in the UV data on a WD in a CV was made with the observation of U Gem 
using the Hopkins Ultraviolet Telescope \cite{Long_ugem93}.  The spectrum obtained 
about 11 days after U Gem had returned to quiescence from a normal outburst
is shown in Figure 1.  The spectrum shows not only several members of the Lyman series in absorption, but also a significant number of narrow absorption lines 
expected from a hot WD atmosphere salted with metal-rich material from an accretion disk.  The overall spectrum is consistent with a WD surface temperature of 
38,000 K and normal abundances, although Long et al.\ [1993] argued that a better
fit to the data could be obtained if the 85\% of the WD surface were 30,000 K and 
15\% were 57,000 K.  Long et al.\ associated the higher temperature region with an
accretion belt left over from the previous outburst, and suggested that the cooling
of the accretion belt accounted for the decline in UV flux which U Gem exhibits
through the interoutburst interval (and which was known from IUE observations \cite{KSS}).  

\begin{figure}
\centering
\begin{minipage}[c]{0.40\textwidth}
\centering
\includegraphics[width=0.9\textwidth,angle=270]{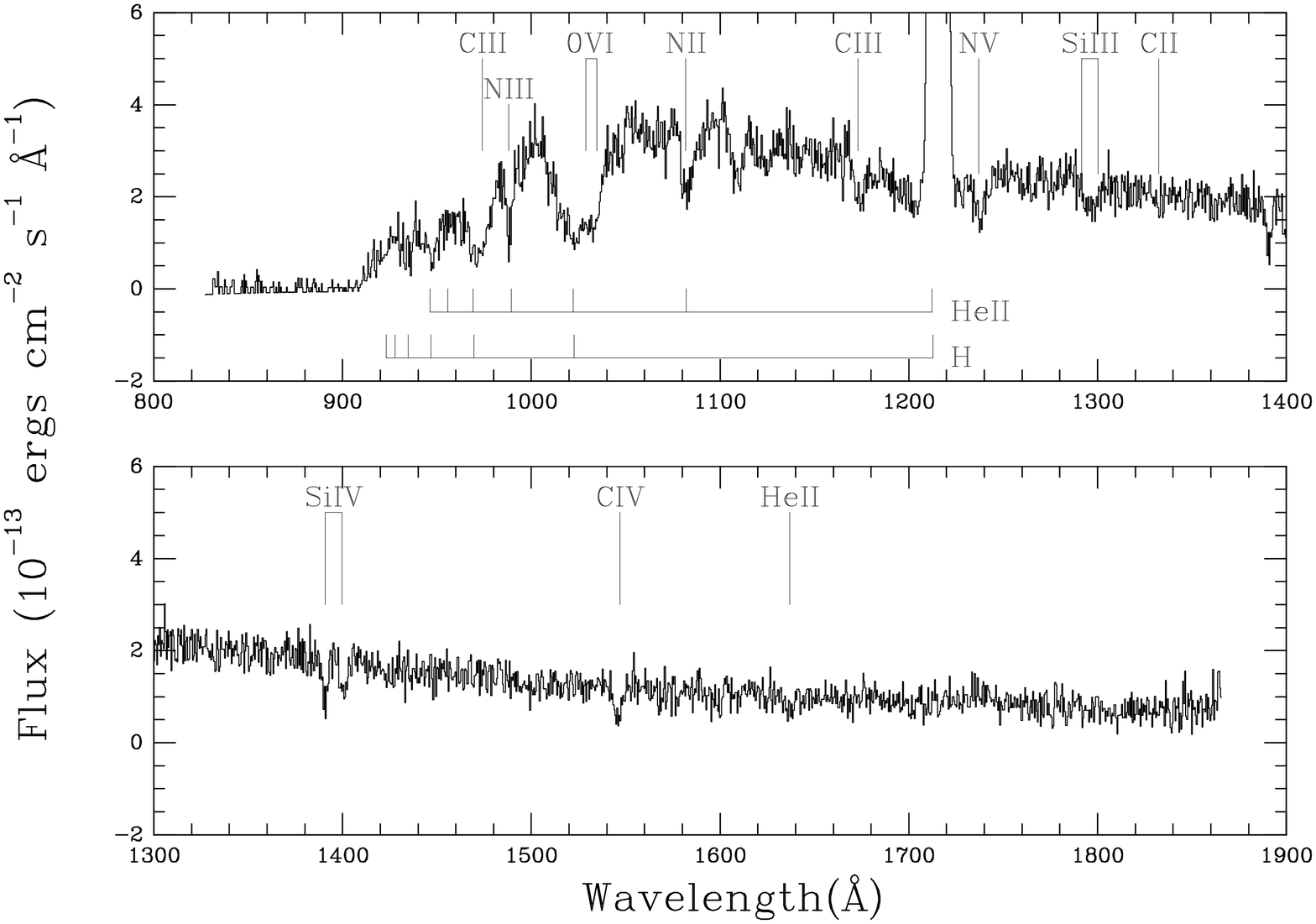}
\caption{The HUT spectrum of the WD in U Gem shortly after return to quiescence. }
\end{minipage}
\hspace{0.1\textwidth}
\begin{minipage}[c]{0.40\textwidth}
\centering
\includegraphics[width=0.9\textwidth,angle=270]{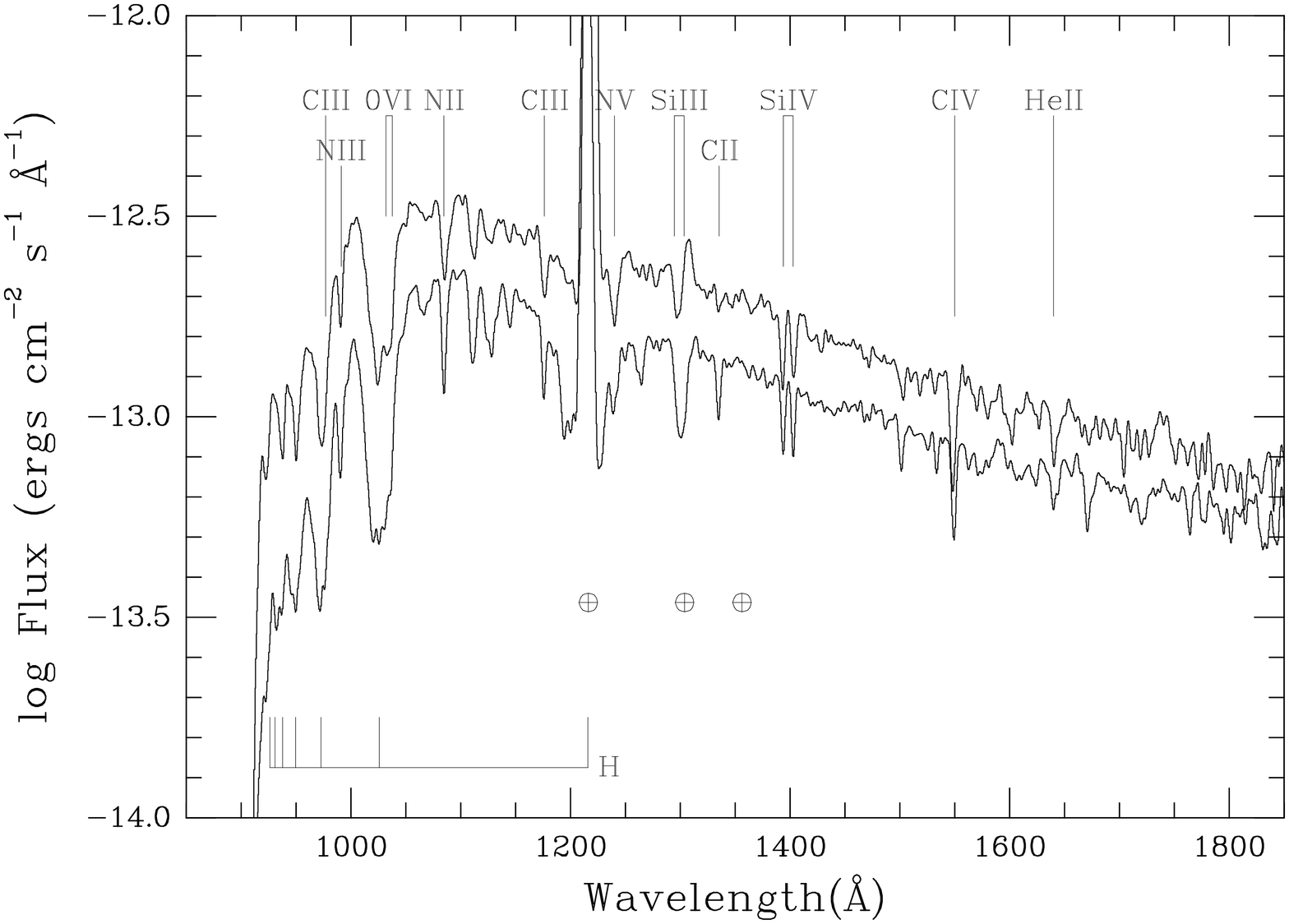}
\caption{A comparison between the spectra of U Gem obtained on Astro 1 and Astro 2.  The difference in flux reflects that fact that the second observation occurred 185 days into the outburst cycle.}
\end{minipage}
\end{figure}

Subsequent observations with HST and HUT have confirmed both that the effective 
temperature of U Gem, as observed in the FUV, drops through early portion of 
the interoutburst period and that flux decline is less than expected if the 
entire WD cools.  For example, Long et al.\ [1994], using the FOS on HST, found that 
the mean temperature of the WD dropped from 39,400 K 11 days after an outburst 
to 32,100 K 70 days after the same outburst.   Figure 2 shows a comparison between
an observation with HUT on Astro 2 in 1994 of U Gem 185 days from an outburst 
and the spectrum obtained on Astro-1 in 1990 11 days from outburst \cite{Long_ugem95}.  
The difference
in temperature is evident in the relative fluxes at 950 \AA\ compared to 1450 \AA, the depth of the Lyman lines, and the specific ions in the spectrum.

The nature of the ``second source'' in U Gem is not resolved.  It is not known 
whether it is a result of a structure left over on the WD surface from the 
previous outburst, or whether it is a result of ongoing accretion after the outburst.
At 1400 \AA, the flux from the ``second source'' could be produced by an optically thick accretion disk with an accretion rate of \EXPU{<1}{15}{g~s^{-1}}.  
Meyer \& Meyer-Hofmeister [1994] have suggested the accretion rate in quiescence gradually declines after an outburst as the inner accretion disk is eaten away.  

It is also not known how best to model the second source.  Long et al.\
[1993; 1995] used a non-rotating WD; Cheng et al.\ [1997] considered an ``accretion ring''.  
The difficulties are that higher S/N HST observations can only be carried out 
longward of 1150 \AA\ where the ``second source'' fraction of emission is small, 
that it is not very clear what the spectrum of an accretion ring should be, and 
that the number of parameters can be large if abundances are allowed to vary (as Cheng et al.\ assumed).  The best hope to resolve these difficulties is likely an intensive survey of U Gem with HST, or possibly FUSE, with multiple observations in a single outburst interval.
 
\begin{figure}
\centering
\begin{minipage}[c]{0.40\textwidth}
\centering
\includegraphics[width=0.9\textwidth]{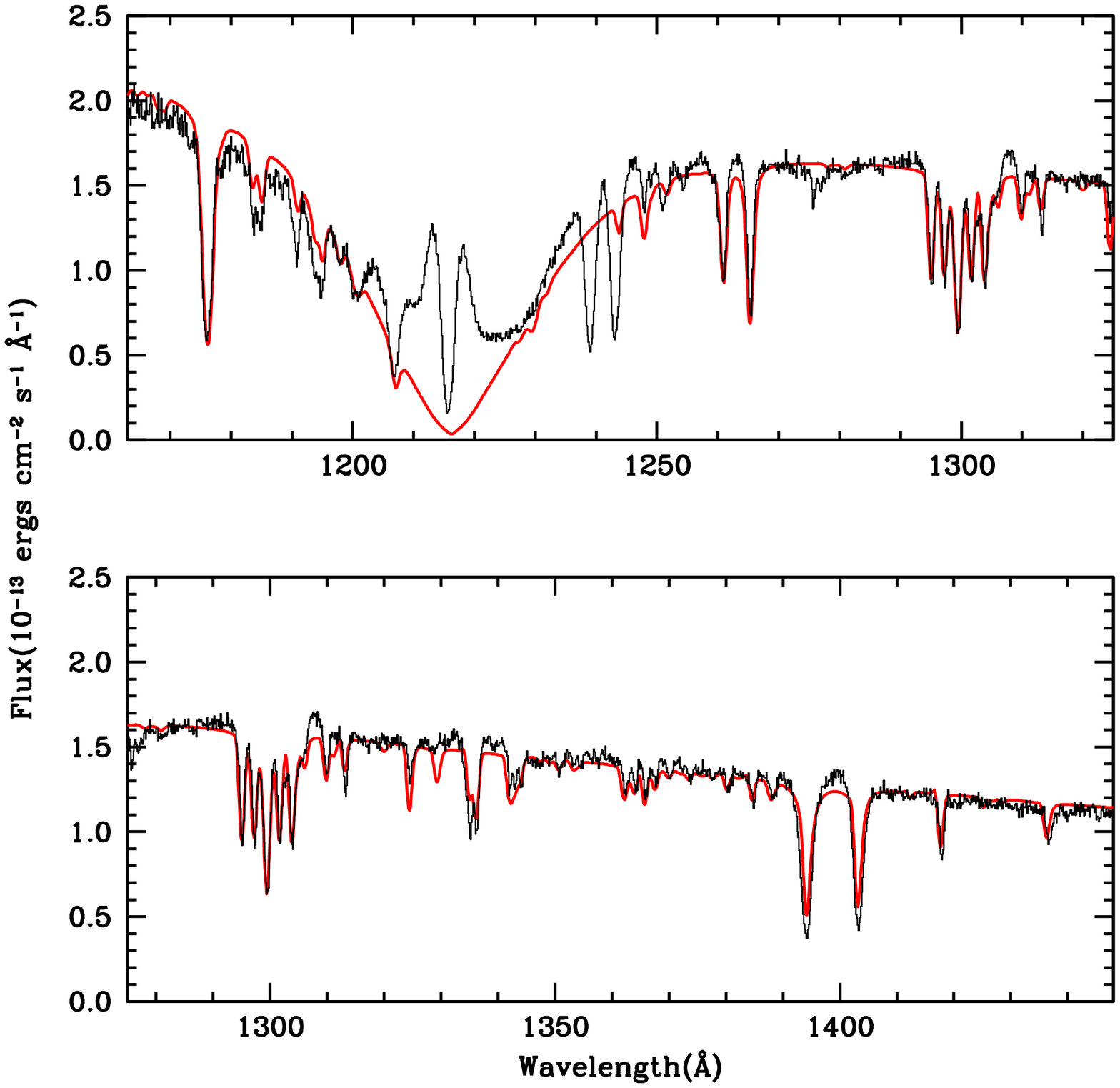}
\caption{A comparison between a high S/N spectrum (black) of U Gem in quiescence to a model WD spectrum.  Sub-solar C abundances and enhanced N are required.}
\end{minipage}
\hspace{0.1\textwidth}
\begin{minipage}[c]{0.40\textwidth}
\centering
\includegraphics[width=0.9\textwidth]{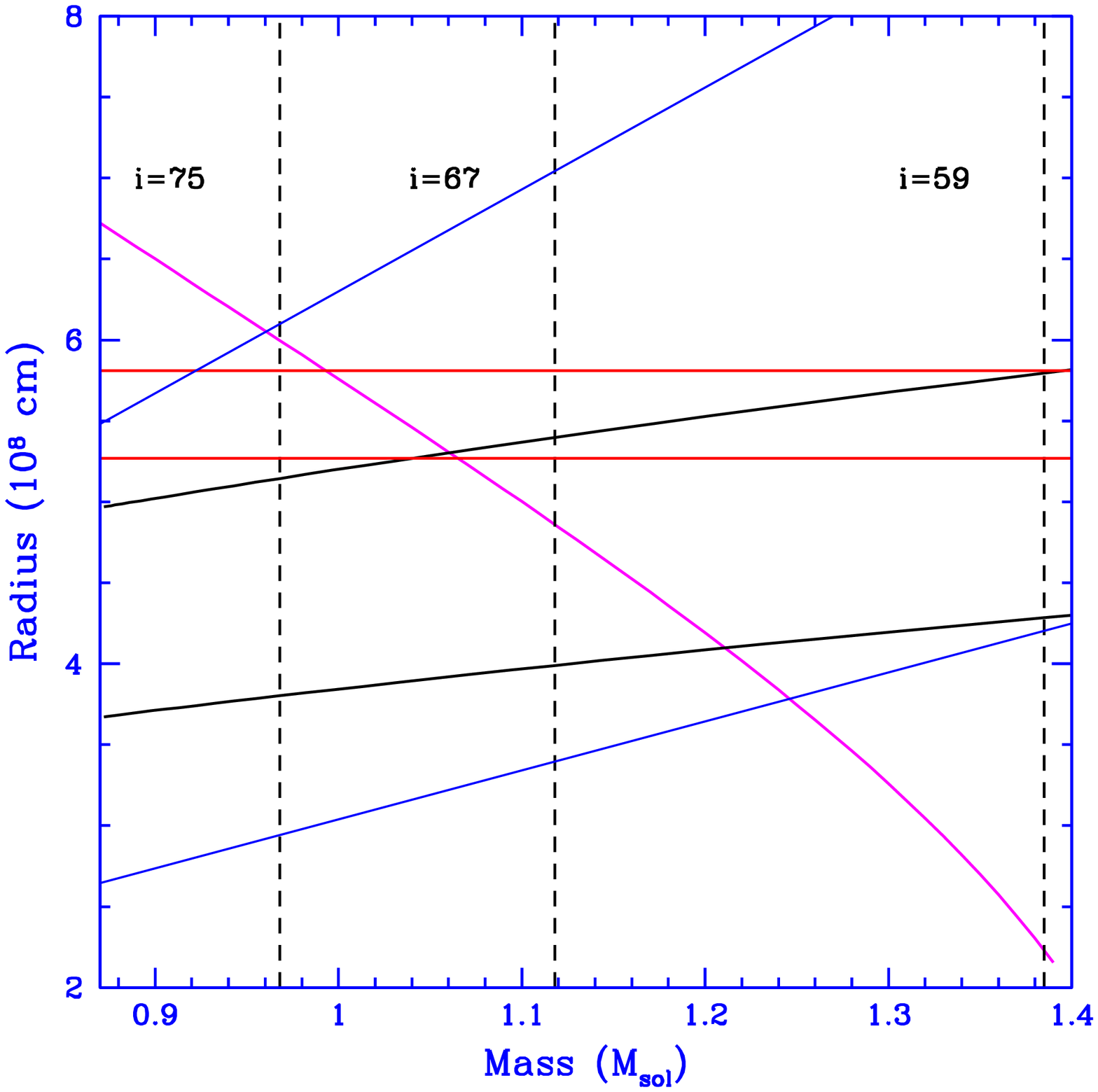}
\caption{Constraints on the mass and radius of the WD in U Gem based on recent analyses of the UV spectrum of the WD and the astrometric distance.}
\end{minipage}
\end{figure}

The HUT and early HST studies of U Gem assumed normal abundances.  However, independent analyses of two different high S/N GHRS datasets of U Gem far from outburst 
show clear evidence of CNO processing in the material in the photosphere of U Gem.
Sion et al.\ [1998] found C to be 0.05$\times$ solar, N 4$\times$ solar and Si 0.4$\times$ solar.  
Similarly, Long \& Gilliland [1999] found  C to be 0.1$\times$ solar, N 4$\times$ solar, Si 0.4$\times$ solar, 
and Al 0.4$\times$ solar.  A comparison between the model obtained by Long \& Gilliland and the data is shown in Figure 3.  
At this point,  systematic errors in the calibration of GHRS data, errors due to the ``second source'' in the actual model spectra, 
and atomic parameters dominate the errors.  The N abundance is really due to one
line complex at 1184 \AA.  
(Note however that there are additional N features in the HUT spectra and these too indicate a significant overabundance of N relative to C.)

These and other high S/N spectra have also been used to derive very accurate 
measurements of the rotation velocity ($<100 \VEL$ \cite{Sion_ugem94}, $\gamma$ velocity,
and K$_1$ velocity of the WD in U Gem.  
The low rotation velocity may account for the fact that for U Gem in outburst
the boundary layer and disk luminosity are comparable, unlike  SS Cyg and VW Hyi \cite{Long_euve}.
Figure 4 is a graphical representation of the current constraints on the mass and radius of U Gem, an update of a figure presented by Long \& Gilliland [1999] to reflect the astrometric 
distance determination for U Gem \cite{harrison99}.  The solid blue lines bound the
allowed region established by the gravitational redshift of the WD; this region is
broad primarily because of the uncertainty in the $\gamma$ velocity of the secondary.The solid black curves bound the region allowed by UV measurements of the temperature and flux and the Bailey relation for the distance.  The red lines bound the region
allowed by the preferred astrometric distance and the temperature and flux.  The 
magenta curve represents a standard WD mass-radius relation.  
Assuming this applies,
the mass of the WD in U Gem is $1.02\pm0.04$ \MSOL and the radius is 
\EXPU{5.6\pm0.2}{8}{cm}.  The inclination for U Gem is $\sim70^{o}$, somewhat greater than usually assumed.

\begin{figure}
\centering
\begin{minipage}[c]{0.40\textwidth}
\centering
\includegraphics[width=0.9\textwidth,angle=270]{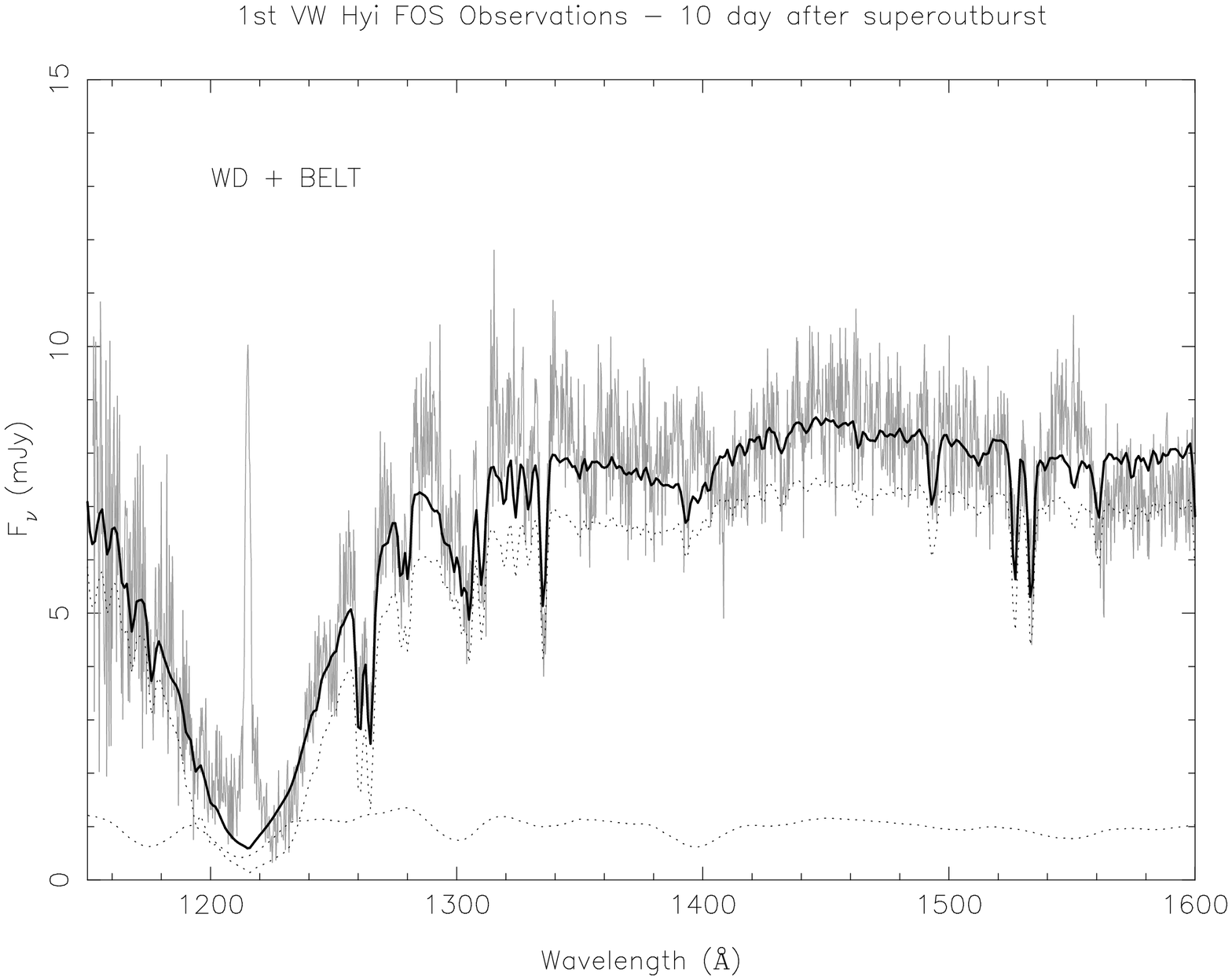}
\caption{A comparison between the spectrum of VW Hyi in quiescence as observed  VW Hyi with the FOS compared to a model consisting of a WD and an accretion disk ring.}
\end{minipage}
\hspace{0.1\textwidth}
\begin{minipage}[c]{0.40\textwidth}
\centering
\includegraphics[width=0.9\textwidth,angle=270]{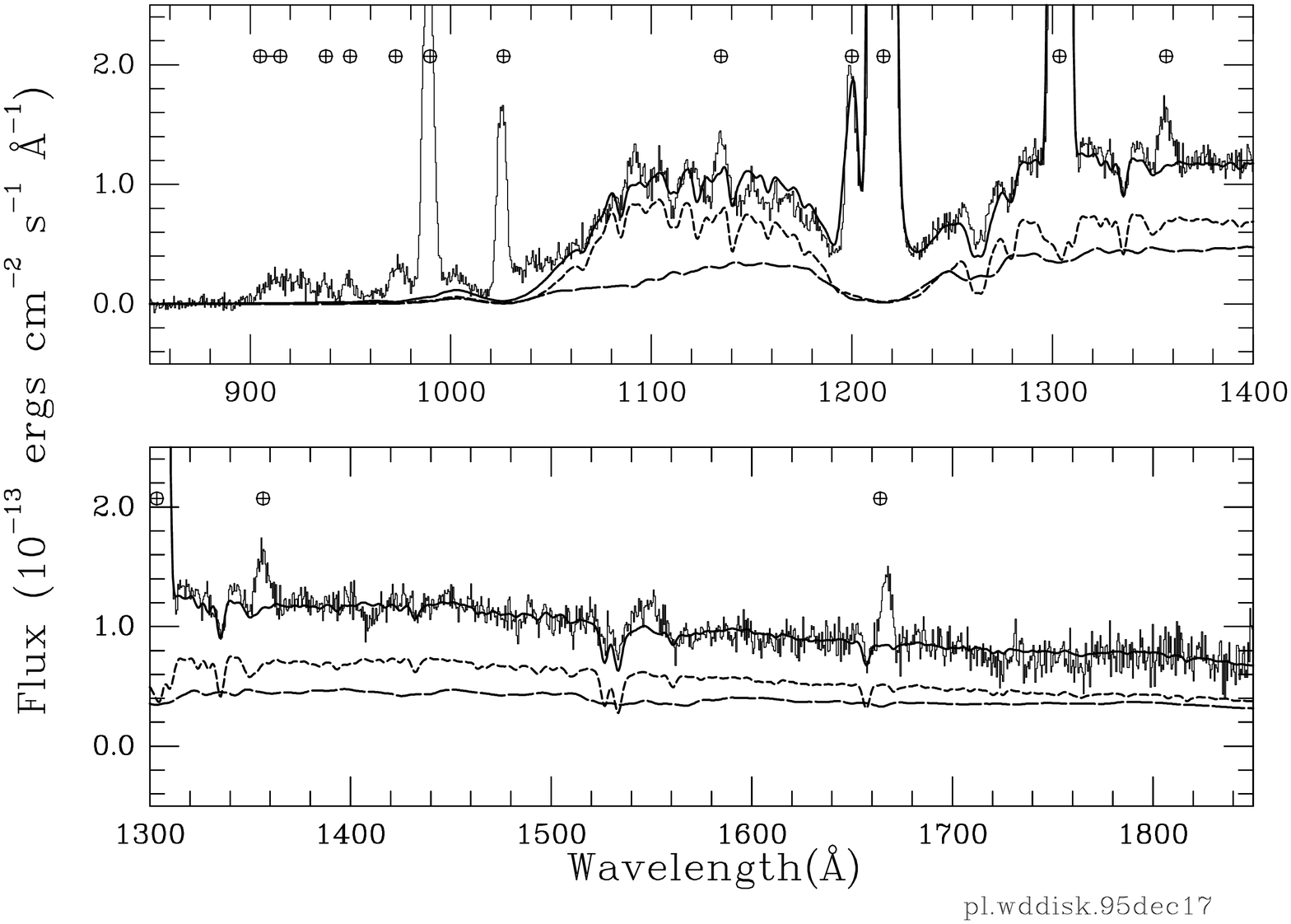}
\caption{The HUT spectrum of VW Hyi in quiescence compared to a model with contributions from WD and an optically thick accretion disk.}
\end{minipage}
\end{figure}

\section{VW Hydri}

Like U Gem, VW Hyi has been the subject of a variety of UV studies using instrumentation on HST and HUT. Like U Gem, there appears to be at least two sources of 
UV emission. In VW Hyi, the ``second source'' is somewhat 
more prominent.  Figures 5 and 6 show comparisons between models and FOS and HUT
data, respectively.  Sion et al [1996] compared the FOS data to a WD or alternatively
a WD and an accretion belt 
(by which they mean a rapidly rotating ring which radiates like a rotating stellar atmosphere).  
They found that WD had a temperature of 22,500 K, N:C ratios which suggest CNO processing, and a v sin(i) of 300 $\VEL$.  
They found the ring had a somewhat greater temperature 26,000 K and v sin(i) of 3350 $\VEL$, close to that expected from Keplerian motion. 
Long et al.\ [1996b] favored a model consisting of a relatively normal 
abundance WD with
a temperature of 18,700 K and
an optically thick accretion disk with an accretion rate of \EXPU{2.4}{15}{g~s^{-1}}.

Sion et al.\ [1997] have also obtained the GHRS spectrum of VW Hyi shown in 
Figure 7.  
These data were taken 30 days after a normal outburst.  
The data show a WD with v sin(i) of 400 $\VEL$, K$_{1}$ of 69 $\VEL$ and a gravitational redshift of 25-61 $\VEL$.  
The implied mass is 0.86 \MSOL and the radius is \EXPU{6.5}{8}{cm}.  Thus in VW Hyi, as in U Gem, the physical parameters are tightly constrained.

However, the most intriguing feature in the GHRS spectrum is the absorption line 
near 1250 \AA.
Sion et al., after considering several alternatives, conclude that the line is most likely due to PII, and that P abundance $\sim900\times$ solar is required.  
If correct, this would provide the very strong evidence that material in the photosphere of VW Hyi has been processed in the thermonuclear runaway of a nova explosion. 
Since the diffusion time scale on the surface of a WD is rather short, this material would 
have to be maintained in the photosphere by some kind of mixing process or by accretion from the secondary star.
This observational claim is significant and requires strong verification.  
There are features in the spectra of some 
B3 stars observed with IUE, e.g. $\eta$ UMa, which,  if convolved to simulate a 
system with v sin (i) of VW Hyi, look rather like what is observed in the VW Hyi 
spectrum.  However, it is not clear these would be expected in the metal-enriched 
photosphere of a WD.  If the abundance of P is high, then 
it should be possible to verify the result with additional observations 
since there are other features of PII of comparable strength which should appear at 
other UV wavelengths.

\begin{figure}
\centering
\begin{minipage}[c]{0.40\textwidth}
\centering
\includegraphics[width=0.9\textwidth,angle=270]{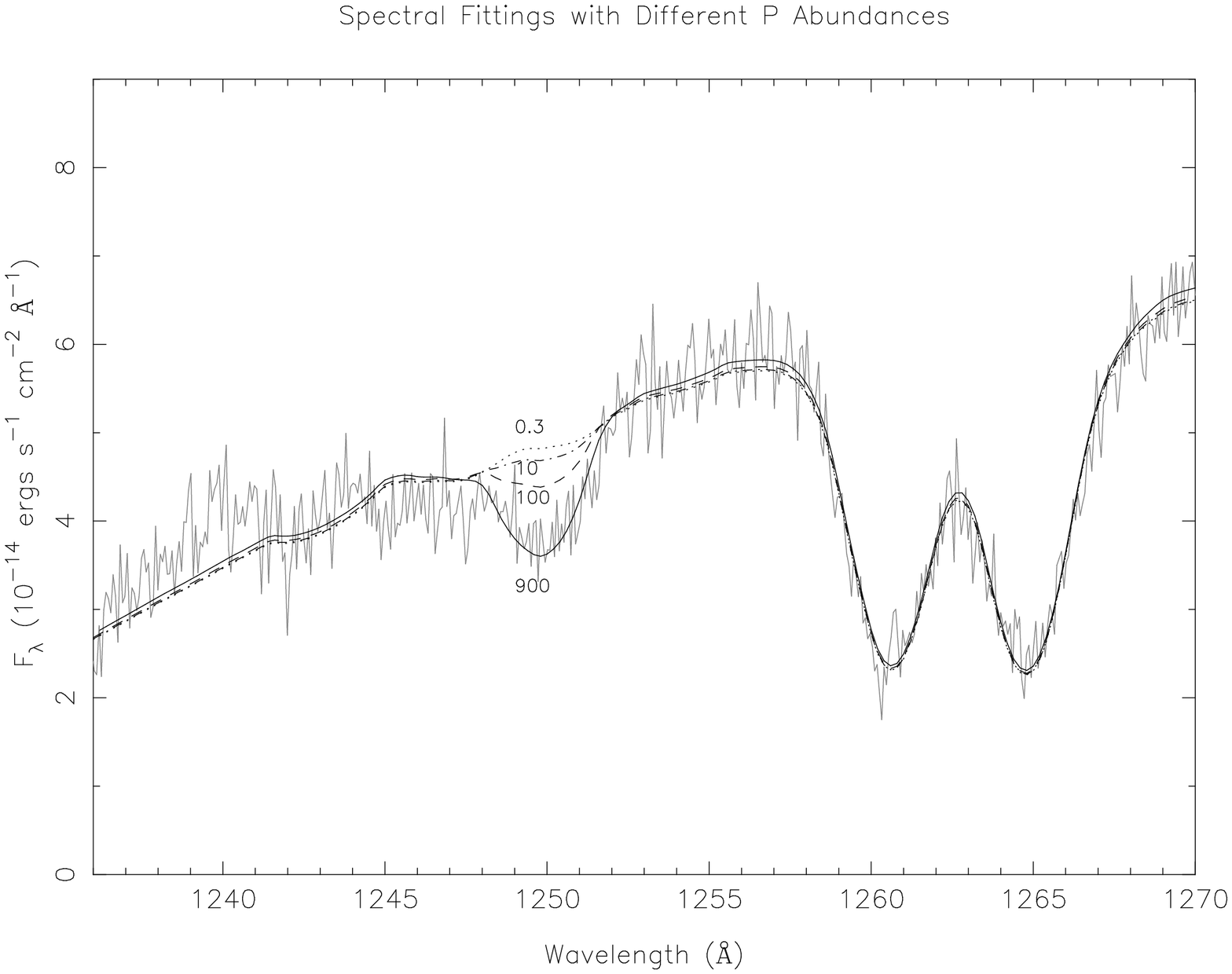}
\caption{A spectrum of VW Hyi in quiescence as observed with the GHRS compared to WD models with varying abundances of P to account for the absorption feature at 1250 \AA.}
\end{minipage}
\hspace{0.1\textwidth}
\begin{minipage}[c]{0.40\textwidth}
\centering
\includegraphics[width=0.9\textwidth]{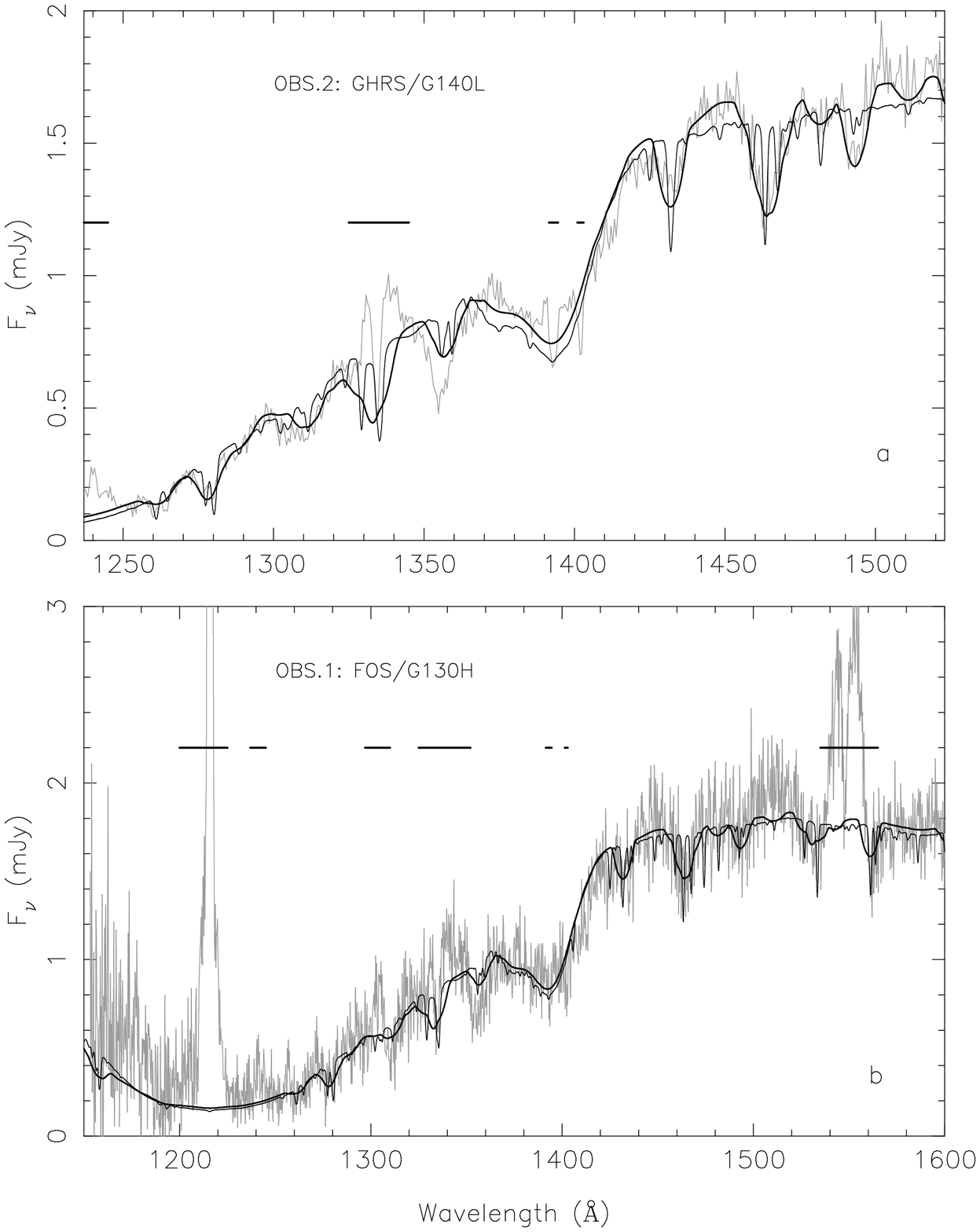}
\caption{FOS spectra of WZ Sge compared to non-rotating and rotating WD models.} 
\end{minipage}
\end{figure}

\section{WZ Sagittae}

The third relatively clean WD in a DN which has been studied with HST 
is WZ Sge \cite{Cheng97b}.  This system is interesting because it has the largest outburst 
amplitude (7 m), the longest interoutburst period (33 years), and of the shortest orbital periods (81 m) of any DN.  
The spectrum indicates a temperature of 14,800 K, and v sin(i) of 1200 $\VEL$ and a periodicity of $\sim$28 s, a periodicity that had been observed previously at optical wavelengths.  The fact that the periodicity is observed in the UV suggests to Welsh et al.\ [1997] that the WD in WZ Sge is magnetic.  
The large rotational velocity implies that radius of the 
WD in WZ Sge would be expected to be significantly greater than 
for a non-rotating WD of the same mass.  
From the value of v sin(i), a periodicity in the data near 28 s, and 
estimates of the inclination of the system, Cheng et al.\ estimate a WD radius 
of \EXPU{5.5}{8}{cm}.  From the spectral 
analysis, log(g) appears to be $\sim8.0$ and in that case $M_{WD}\sim0.3$ \MSOL, 
much lower than VW Hyi or U Gem.
The large value of v sin (i) may complicate the  abundance analyses, 
but the indication in WZ Sge is that C is more abundant than N.   

\section{Summary and Future Prospects}

The decade of the 90s has been the decade of the first quantitative spectroscopy 
of WDs.  For U Gem, VW Hyi, and WZ Sge, the observations are in fact sufficient 
to determine the basic physical properties of WDS and to begin to challenge our understanding of the physics of WDs in DNe.  

Much remains to be done however.  Observers
need to study more systems at high S/N, with good resolution, 
and covering simultaneously a large wavelength range to disentangle the 
various sources in the system.  Multiple observations especially in a single 
outburst interval are needed 
to understand the response of the WD to the outburst and ongoing accretion.  
And accurate parallaxes are need to tie down the distances.  Theorists
need to develop a better understanding of the structure quiescent disks so that
modelers will be able to calculate the spectra of this contaminating component of the the WD spectrum.  And modelers need to attempt to calculate the 
uncertainties in their modeling more accurately, comparing stars we understand 
with the few WDs we are able to observe in CVs with more normal stars.  A new millennium is coming and so we all have the right to be optimistic.

\end{document}